\documentclass[aps,superscriptaddress,nofootinbib,preprint]{revtex4}

\usepackage{amsmath}
\usepackage{graphicx}
\usepackage{color}

\begin{document}

\begin{flushright}
CYCU-HEP-14-11
\end{flushright}

\title{n-partite information in Hawking radiation}

\author{Shogo Kuwakino}\thanks{%
E-mail: shogo.kuwakino@gmail.com}
\affiliation{Department of Physics and Center for High Energy Physics, Chung Yuan Christian University, Chung Li City, Taiwan}

\author{Wen-Yu Wen}\thanks{%
E-mail: steve.wen@gmail.com}
\affiliation{Department of Physics and Center for High Energy Physics, Chung Yuan Christian University, Chung Li City, Taiwan}
\affiliation{Leung Center for Cosmology and Particle Astrophysics\\
National Taiwan University, Taipei 106, Taiwan}

\begin{abstract}
We study the entanglement among sequential Hawking radiations in the Parikh-Wilczek tunneling model of Schwarzschild black hole.  We identify the part of classical correlation and that of quantum entanglement in bipartite information and point out its imitated relation to quantum gravity correction.  Explicit computation of n-partite information shows that it is positive (negative) for even (odd) $n$, which happens to agree with the holographic computation. The fact that entanglement in the mutual information grows with time mimics the second law of thermodynamics.  Later we extend our study to the AdS black hole and find the total mutual information which includes the classical correlation is sensible to the Hawking-Page phase transition.
\end{abstract}

\maketitle



\section{Parikh-Wilczek tunneling model with quantum correction}

The Parikh-Wilczek model \cite{Parikh:1999mf} gives us an intuitive description of black hole evaporation, that is, Hawking radiation \cite{Hawking:1974rv, Hawking:1974sw} from the Schwarzschild black hole is described by a tunneling process across the black hole horizon \cite{Braunstein:2011gz}. The semiclassical Hawking temperature is derived by evaluating semiclassical emission rate, and it can be written in terms of the difference of entropies before and after the emissions. In the case of a radiation from black hole, which we denote a quanta as $\omega$ and the black hole mass as $M$, the tunneling rate is given by 
\begin{eqnarray}
&&\Gamma \sim e^{\Delta S},
\end{eqnarray}
where $\Delta S$ is the change in the black hole entropy during the emission process
\begin{eqnarray}
&&\Delta S = -8\pi M  \omega + 4 \pi \omega^2.
\end{eqnarray}
In \cite{Medved:2005vw, Banerjee:2008ry}, after quantum correction to surface gravity is included, the logarithmic correction to the Bekenstein-Hawking area law for a sufficiently large black hole mass $M$ is derived as
\begin{equation}
S(M) = 4\pi M^2 - 8\pi \alpha \ln{M}.
\end{equation}
Here $\alpha$ is a model dependent dimensionless parameter, and the sign of $\alpha$ could be either positive or negative depending on with which spin particles are included in the loop calculation.

In this letter, after reviewing mutual information in the framework of the Parikh-Wilczek model, we discuss its generalization to n-partite information. We show that a behavior of n-partite information, positive (negative) for even (odd) $n$, agrees with the holographic computation. We also show that, in the case of AdS Schwarzschild black hole, the quantity of total mutual information can be a reference index for the Hawking-Page phase transition.

\section{Mutual information and conservation of entropy}

Thanks to the nonthermal spectrum in the Parikh-Wilczek model, one expects some kind of entanglement or correlation between two successive emissions.  There is no clear picture of which microstates responsible for black hole entropy, not to mention entanglement among them.  However, it is the academical guess to imagine such entanglement is via quantum gravity and its quantity could be accumulated and classically described by some extensive macroscopic quantity, such as black hole mass and entropy \cite{Azeyanagi:2007bj}.  Entropy loss for a black hole of mass $M$ due to a radiation of a mass quantum $\omega_i$ reads
\begin{equation}
\label{eqn:entropy_change}
S_E(M,\omega_i) \equiv S(M) - S(M-\omega_i)  = 8 \pi (M-\omega_i)\omega_i + 4\pi \omega_i^2  + 8\pi \alpha \ln(1-\frac{\omega_i}{M}).
\end{equation}

Since conservation of entropy is one essential feature in the Parikh-Wilczek tunneling model, the entanglement between the black hole and the radiation, if any, should be able to identify in (\ref{eqn:entropy_change}).  The first two terms are {\sl classical} for its survival after quantum correction is removed for $\alpha \to 0$.  In particular, we would like to call the first term as the {\sl classical correlation} between the black hole and the emission quanta, which is proportional to total Newton's gravitational force flux between two massive objects.  The second term is seemed as the {\sl self correlation}, which measures the maximum entropy a radiation could carry as if it collapsed into a small black hole.  This leaves us the last term as a candidate for the desired quantum entanglement.  Though there is only single emitted quantum involved, this term actually represents the entanglement between the black hole and one emission, denoting as
\begin{equation}\label{eqn:one_partite}
I^{[1]}_{i} = 8\pi \alpha \ln(1-\frac{\omega_i}{M}).
\end{equation}
The entanglement is negative in generic and becomes $0$ at the zero temperature limit $M\to \infty$.
Then the bipartite information among two emitted quanta $\omega_i$ and $\omega_j$ can be defined by the conventional way:
\begin{equation}\label{eqn:bipartite_quantum}
I^{[2]}_{ij}\equiv I^{[1]}_{i} + I^{[1]}_{j} - I^{[1]}_{(ij)} =  8\pi \alpha \ln{\frac{(M-\omega_i)(M-\omega_j)}{M(M-\omega_i-\omega_j)}}.
\end{equation}

Since the bipartite information is expected to be positive definite, this suggests $\alpha >0$\footnote{The positive $\alpha$ is also suggested in several discussions, such as effects of loop corrections \cite{Fursaev:1994te} and a black hole remnant \cite{Xiang:2006mg, Chen:2009ut}}.
On the other hand, if we define another quantity, denoting as {\sl mutual information}, to measure how does the entropy change for an emission $\omega_j$ depend on whether an earlier emission $\omega_i$ happens or not, that is 
\begin{equation}\label{eqn:mutualinfo}
E^{[2]}_{ij} \equiv S_E(M,\omega_j \big|\omega_i)-S_E(M,\omega_j)= 8\pi \omega_i\omega_j + I^{[2]}_{ij},
\end{equation}
where $S_E(M,\omega_j \big|\omega_i) = S_E(M-\omega_i,\omega_j)$ for the tunneling model \cite{Zhang:2009jn}.  We see that $I^{[2]}$ exactly captures the quantum part of this mutual information.  The difference is the classical correlation among two emitted quanta.   We remark that if it were for thermal Hawking radiation, one expects two emissions are totally independent event or mass loss due to previous emission can be ignored.  To see this, we could first rescale $\omega_i \to \epsilon\omega_i$ and equations (\ref{eqn:bipartite_quantum}) and (\ref{eqn:mutualinfo}) become
\begin{eqnarray}
&&I^{[2]}_{ij} = 8 \pi \alpha \ln{\frac{M^2-\epsilon(\omega_i+\omega_j)M+\epsilon^2 \omega_i\omega_j}{M^2-\epsilon(\omega_i+\omega_j)M}}, \nonumber\\
&&E^{[2]}_{ij} = 8\pi \epsilon^2 \omega_i\omega_j + I^{[2]}_{ij}.
\end{eqnarray}
It is easy to see that both quantities vanish if we only keep terms up to linear order ${\cal O}(\epsilon)$, even the quantum correction coefficient $\alpha \neq 0$.

\section{n-partite information}

It is straightforward to compute entanglement for $n$ quanta of radiations.  For instance, the tripartite information can be expressed in terms of bipartite information:
\begin{equation}
I^{[3]}_{ijk}\equiv I^{[2]}_{ij}+I^{[2]}_{ik}-I^{[2]}_{i(jk)}=8\pi \alpha \ln{\frac{(M-\omega_i)(M-\omega_j)(M-\omega_k)(M-\omega_i-\omega_j-\omega_k)}{M(M-\omega_i-\omega_j)(M-\omega_i-\omega_k)(M-\omega_j-\omega_k)}}.
\end{equation}
It can be shown explicitly that the subadditive inequality  $I^{[3]}_{ijk}<0$ is satisfied for $0<\omega_i+\omega_j+\omega_k<M$.  It worths to mention that one obtains the same result for $I^{[3]}$ by using the definition (\ref{eqn:mutualinfo}) for bipartite information.  In fact, the term associated with classical correlation always vanishes in the $I^{[n]}$ for any $n>2$.  We believe that this is because the law of Newton's gravitational force is designed for pairs of massive objects, and there does not exist force formula for three or more objects.  

After all, the $n$-partite information can be computed iteratively from $(n-1)$-partite:
\begin{eqnarray}
I^{[n]}_{i_1i_2\cdots i_{n-1}} &&= I^{[n-1]}_{i_1i_2\cdots i_{n-2}i_{n-1}}+I^{[n-1]}_{i_1i_2\cdots i_{n-2}i_{n}}-I^{[n-1]}_{i_1i_2\cdots (i_{n-1} i_{n})}\nonumber\\
&&=8\pi\alpha \ln{\frac{ {\displaystyle \prod_{p} } (M-\omega_{i_p}) { \displaystyle \prod_{p<s<t} }(M-\omega_{i_p}-\omega_{i_s}-\omega_{i_t})\cdots}{M { \displaystyle \prod_{p<s} }(M-\omega_{i_p}-\omega_{i_s}) { \displaystyle \prod_{p<s<t<u} } (M-\omega_{i_p}-\omega_{i_s}-\omega_{i_t}-\omega_{i_u})\cdots}},
\end{eqnarray}
where indexes $p, s, t, u, \cdots$ run through $1, 2, \cdots, n$.
Note that there is always odd (even) number of $\omega_i$ in each product term in the numerator (denominator).

In the Appendix \ref{app1}, we prove that the $n$-partite information is always positive for even $n$ and negative for odd $n$.  The alternative sign is in agreement with recent computation in holographic spatial partite \cite{Alishahiha:2014jxa}.  

\section{Total bipartite information and arrow of time}

Here we define another quantity which is a generalization of mutual information, denoting as total bipartite information, defined by
\begin{equation}
J^{[n]}_{i_1 i_2\cdots i_n}  \equiv \sum_{k=1}^n  I^{[1]}_{i_k} - I^{[1]}_{i_1 i_2 \cdots i_n  },
\end{equation}
or equivalently, this quantity can be described by summing up all bipartite information among $n$ emissions:
\begin{equation} \label{FunJ}
J^{[n]}_{i_1 i_2\cdots i_n}  = I^{[2]}_{i_1i_2} + I^{[2]}_{(i_1i_2) i_3} + I^{[2]}_{(i_1 i_2 i_3) i_4} + \cdots + I^{[2]}_{(i_1 i_2 \cdots i_{n-1}) i_n} 
=  8 \pi  \ln \frac{ { \displaystyle \prod_p } (M-\omega_{i_p})}{M^{n-1}(M- { \displaystyle \sum_p } \omega_{i_p})}.
\end{equation}
It turns out that $J^{[n]} $ grows with number of emissions $n$ since the bipartite information in \eqref{FunJ} is positive definite. In the Appendix \ref{app2}, we prove that this function is a  decreasing function with increasing black hole mass $M$.  Since we know that $n$ increases or $M$ becomes less at later time in the process of Hawking radiation, this monotonically-growing quantity could be interpreted as quantum entanglement entropy, which plays a similar role as in the second law of thermodynamics.  We remark that the entanglement entropy $J^{[n]}$ could serve as a quantity running with RG flow in $0+1$-dimension since 
\begin{equation}
\beta \frac{\partial J^{[n]}}{\partial \beta} = M\frac{\partial J^{[n]}}{\partial M} < 0,
\end{equation}
where UV (IR) regime refers to the limit $M\to 0$ ($\infty$).

One can also define the total mutual information to include the classical correlation among $n$ emissions:
\begin{equation}
F^{[n]}_{i_1 i_2 \cdots i_n} \equiv E^{[2]}_{i_1i_2} + E^{[2]}_{(i_1i_2)i_3} + E^{[2]}_{(i_1 i_2 i_3) i_4} + \cdots + E^{[2]}_{(i_1 i_2 \cdots i_{n-1}) i_n} = \sum_{p<s} 8\pi \omega_{i_p} \omega_{i_s} + J^{[n]}_{i_1 i_2 \cdots i_n}.
\end{equation}
This quantity also grows with $n$ thanks to increasing pairwise correlation. However we shall see in the next section that it does not always behave in this way.

\section{Total mutual information and Hawking-Page phase transition}

A Schwarzschild black hole has a negative specific heat in the asymptotic flat spacetime so it cannot help but radiate till the last bit.   However, a black hole in the asymptotic AdS behaves differently while the AdS boundary acts like a reflecting wall.  A small black hole still behaves similarly as in the flat spacetime, however a large black hole would reach thermal equilibrium with its own radiation reflected from the boundary wall.  The phase transition happening in between these two limits is called the Hawking-Page phase transition \cite{Hawking:1982dh}.  While the total bipartite information $I^{[n]}$ always grows with $n$, the total mutual information $F^{[n]}$ needs not behave the same.  In the following, one can show $F^{[n]}$ could turn into a decreasing function with $n$ for large enough black hole $M > L^2$.  

Now consider a AdS$_5$ Schwarzschild black hole with horizon at 
\begin{eqnarray}
&& r_+^2 = \frac{L^2}{2} \left( \sqrt{1+ \frac{ 32 M}{ 3 \pi L^2}  }-1 \right).   
\end{eqnarray}
Here $L$ means the AdS radius of curvature, and we set the Newton's constant in five dimensions to be unity. 
Without loss of generality, one can simply investigate the case of $n$ emitted quanta of same mass $\omega$, that is
\begin{eqnarray}
F^{[n]}(M,\omega,\cdots,\omega) &=& n S_E(M,\omega) - S_E(M, n\omega) \nonumber \\ 
&=& \frac{ \pi^2 L^3}{2^{5/2}} \left(  C^{[n]} (\omega)+ J^{[n]}(\omega)  \right),
\end{eqnarray}
with
\begin{eqnarray}
C^{[n]} & \equiv & n \left[ \left( \sqrt{1+\frac{32 M}{ 3\pi L^2}}-1 \right)^{3/2}
- \left( \sqrt{1+\frac{ 32 (M-\omega) }{ 3 \pi L^2}  } -1 \right)^{3/2} \right]  \nonumber\\
&&- \left[ \left( \sqrt{1+ \frac{ 32 M}{ 3 \pi L^2}   }-1 \right)^{3/2}- \left( \sqrt{1+\frac{ 32 (M-n\omega)}{ 3 \pi L^2}   }-1  \right)^{3/2}  \right], \\
J^{[n]} & \equiv & \frac{3 c}{2} \ln \frac{ \left( \sqrt{1+ \frac{ 32 (M-\omega)}{ 3\pi L^2}  }-1 \right)^n}{ \left( \sqrt{1+ \frac{ 32 M}{ 3\pi L^2 }  }-1 \right)^{n-1} \left( \sqrt{1+ \frac{ 32 (M-n\omega)}{ 3 \pi L^2}  }-1 \right) }.
\end{eqnarray}
Here we introduced one dimensionless parameter $c$, which value corresponds to an overall coefficient of logarithmic correction term to a black hole entropy formula. The value of $c$ is analyzed in \cite{Das:2001ic, Mukherji:2002de} to be $c=1/2$ or $c=5/6$. However, in the following, we take $c = 1/2$ since the difference of the values is not sensitive to our result because of the small correction.

\begin{figure}[htbp]
  \begin{center}
    \includegraphics[clip,width=14.0cm]{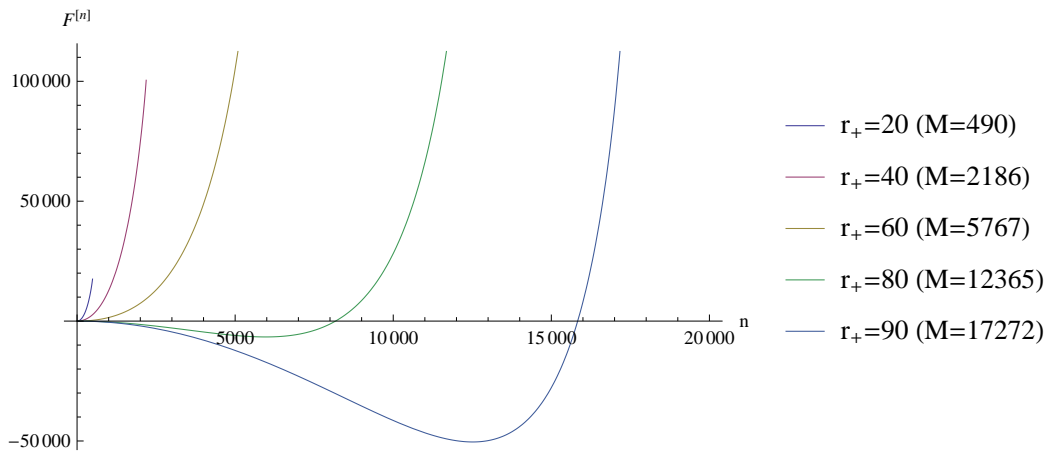}
    \caption{Plots of total bipartite information $F^{[n]}$ against $n$ particle emissions for various values of AdS black hole radius $r_+$ (or AdS black hole mass $M$). Parameters are taken to be $L=100$, $\omega =1 $ and $c=\frac{1}{2}$.}
    \label{fig:Jvsn1}
  \end{center}
\end{figure}

In Figure \ref{fig:Jvsn1}, we plot the total bipartite information $F^{[n]}$ against $n$ particle emissions for various values of AdS black hole radius $r_+$. The AdS black hole mass $M$ can be described in terms of the radius $r_+$ by the expression,
\begin{eqnarray}
&&M = \frac{3 \pi}{8} r_+^2 \left( \frac{r_+^2}{L^2} + 1 \right). 
\end{eqnarray}
In the figure we set the AdS radius $L =100$ and mass of emitted quanta $\omega =1$. 

\begin{figure}[htbp]
  \begin{center}
    \includegraphics[clip,width=10.0cm]{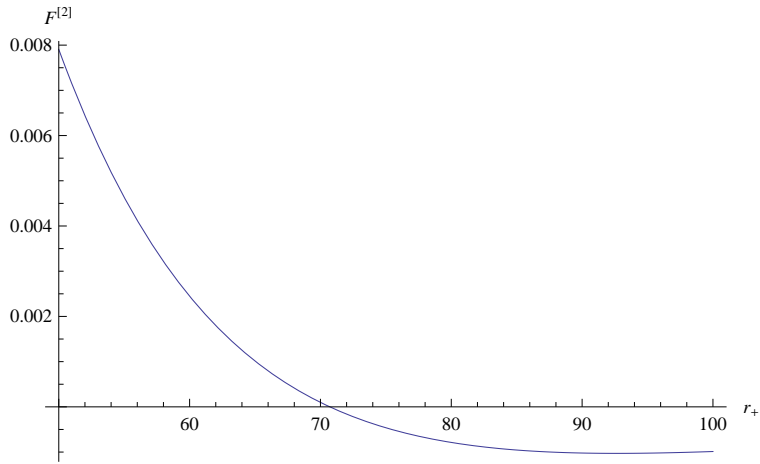}
    \caption{Plots of total bipartite information $F^{[2]}$ against AdS black hole radius $r_+$. $F^{[2]}$ changes the sign at $r_+ \sim 71$ or $M \sim 8932$. Parameters are taken to be $L=100$, $\omega =1 $ and $c= \frac{1}{2}$.}
    \label{fig:Criticalr+}
  \end{center}
\end{figure}

From this figure we find that there exists a critical AdS black hole radius $r_{+cri}$ (or a critical AdS black hole mass $M_{cri}$). In $r_+ < r_{+cri}$, the total bipartite information have positive value for any $n$, however, in the case of relatively large black hole $r_+ > r_{+cri}$, the function can be negative, namely $\partial F/\partial M \le 0$, for some value of emission number $n$. For large $n$ the function becomes positive in any case. The critical values $r_{+cri}$ and $M_{cri}$ can be evaluated from $F^{[n=2]}$ shown in Figure \ref{fig:Criticalr+}. In the figure we take the parameters to be $n=2$, $L=100$ and $\omega =1 $. We find that, at $r_{+cri} \sim 71$, $F^{[n=2]}$ turns to be negative. The critical radius $r_{+cri} \sim 71$ corresponds to $M_{cri} \sim 8932$ in terms of black hole mass. 
We would like to guess it is the thermalization which outbeats the entanglement as AdS black hole increases its mass.  Thermal randomness expresses itself only via a decreasing of classical correlation, therefore $F^{[n]}$ can be sensitive to the Hawking-Page phase transition. 

\section{An explanation from quantum quench model}

So far we have seen some nice properties of $n$-partite information and related quantities $J^{[n]}$ and $F^{[n]}$.  Though the definition of (\ref{eqn:one_partite}) comes quite naturally from the tunneling model of Hawking radiation, it would be more satisfying to understand it from the aspect of black hole microstates or some effective theory.  With that being said, we recall that for a two-dimensional conformal field theory quench system, the late time behavior of the entanglement entropy between two separated half regions grows like \cite{CC1, CC2}
\begin{equation}
S_E \sim c \ln \frac{t}{\epsilon},
\end{equation}
for central charge $c$ and some UV cutoff $\epsilon$.
Here we consider the time scale is determined by the Hawking temperature, or period of Euclidean time, that is $\beta= 8\pi M$, and the UV scale can be set by the Planck mass, say $\epsilon \sim m_{p}$.   A quantum state comes in sudden entangled with a black hole would carry the entanglement entropy $S_E \sim \ln (M/m_{p})$ at later time.  Then it is suggestive the equation (\ref{eqn:one_partite}) computes the {\sl relative} entropy for two different black hole masses:
\begin{equation}
I^{[1]} \sim \ln (M-\omega)/m_{p} - \ln M/m_{p} \sim \ln \frac{M-\omega}{M}.
\end{equation}
Two comments are in order.  First, we remark that the logarithmic divergence occured in the Von Neumann-type entanglement entropy becomes irrelevant in the relative entropy, for the UV part being cancelled.  All the $n$-partite information and mutual information are also expected to be UV finite for similar reason.  Second, the fact that all the $n$-partite information $I^{[n]} \to 0$ at the zero temperature limit reflects that the origin of Hawking radiation is indeed quantum effect.  Zero temperature correponds to a classical limit where no quantum entanglement can be observed.  This result agrees with the thermal-mutual information (TMI) defined in the eternal black hole \cite{Morrison:2012iz}.

\section{A remark on black hole remnant}
The origin of logarithmic correction to the area law could attribute to loop correction to the surface gravity.  If the Hawking temperature is modified as follows \cite{Fursaev:1994te}:
\begin{equation}
T_H = \frac{\kappa}{2\pi} (1+\frac{\alpha}{M^2}),
\end{equation}
then the corresponding modified area law would be \cite{Banerjee:2008ry} 
\begin{equation}
S = 4\pi M^2 - 4\pi m_c^2 - 4\pi \alpha \ln\frac{M^2+\alpha}{m_c^2+\alpha},
\end{equation}
where a constant of integration has been included to renormalize $S\to 0$ as $M\to m_c$ for remnant mass.  One will need to modify the definition of one-partite information (\ref{eqn:one_partite}) by 
\begin{equation}\label{eqn:one_partite_remnant}
I^{[1]}_{i} = 4\pi \alpha \ln\frac{(M-\omega_i)^2+\alpha}{M^2+\alpha},
\end{equation}
and $n-$partite information henceforth.  One can easily check that the alternative sign of $I^{[n]}$ for even or odd $n$ still remains true, despite that $I^{[n]}$ of odd (even) $n$ have lower (upper) bound, such that
\begin{eqnarray}
&&4\pi \alpha \ln\frac{m_c^2+\alpha}{(m_c+\omega_i)^2+\alpha} \le I^{[1]}_i < 0, \nonumber\\
&&0 < I^{[n]}_{ij} \le 4 \pi \alpha \ln\frac{((m_c+\omega_i)^2+\alpha)((m_c+\omega_j)^2+\alpha)}{(m_c^2+\alpha) ((m_c+\omega_i+\omega_j)^2+\alpha)},\nonumber\\
&&\cdots
\end{eqnarray}
Therefore, bounded $n$-partite information in Hawking radiation could be a signature for the existence of black hole remnant.

\begin{acknowledgments}
The authors are grateful to hospitality of Physics Department at the Osaka University and useful discussion with Koji Hashimoto, Feng-Li Lin and Shang-Yu Wu during the YITP workshop "Strings and Fields"  and YQIP2014 at Kyoto University, where part of this idea was emerged.  WYW is grateful to comments and references provided by Li-Yi Hsu in the CYCU department.  This work is supported in parts by the Taiwan's Ministry of Science and Technology (grant No. 102-2112-M-033-003-MY4 and No. 103-2633-M-033-003-) and the National Center for Theoretical Science. 
\end{acknowledgments}

\appendix

\section{Proof for sign of $I^{[n]}$} \label{app1}

We will ignore the classical correlation term for $n=1$ and $2$.  For our convenience, we denote $I^{[n]}$ as
\begin{eqnarray}
&&I^{[n]} = 8\pi \alpha \ln{ {I^{[n]}}' },
\end{eqnarray}
where
\begin{eqnarray}\label{eqn:AB_form}
&&{I^{[n]}}' \equiv \frac{A^{[n]}}{B^{[n]}}, \\
&& A^{[n]} \equiv  \prod_p (M-\omega_{i_p}) \prod_{p<s<t}(M-\omega_{i_p}-\omega_{i_s}-\omega_{i_t}) \cdots, \\
&& B^{[n]} \equiv M \prod_{p<s}(M-\omega_{i_p}-\omega_{i_s}) 
\prod_{p<s<t<u}(M-\omega_{i_p}-\omega_{i_s}-\omega_{i_t}-\omega_{i_u}) \cdots.
\end{eqnarray}
We remark that there is always odd (even) number of $\omega_i$ in each product term in the function $A^{[n]}$ ($B^{[n]}$).
It is easy to find that there is a relation between ${I^{[n]}}'$ and ${I^{[n+1]}}'$ as 
\begin{eqnarray} \label{Relation}
&& {I^{[n+1]}}' = \frac{  {I^{[n]}}'  }{  {{I}^{[n]}}' \vert_{M \to M - \omega_{i_{n+1}}}   },
\end{eqnarray}
since the functions $A^{[n]}$ and $B^{[n]}$ satisfy the following relations
\begin{eqnarray}
&& A^{[n+1]} = A^{[n]} \cdot B^{[n]} \vert_{M \to M - \omega_{i_{n+1}}}, \\
&& B^{[n+1]} = B^{[n]} \cdot A^{[n]} \vert_{M \to M - \omega_{i_{n+1}}}.
\end{eqnarray}
In addition, we see that
\begin{eqnarray}
&& \frac{\partial {I^{[n]}}' }{\partial M} 
 = \frac{A^{[n]}}{B^{[n]}} \left(  \frac{1}{A^{[n]}} \frac{\partial A^{[n]} }{\partial M} 
- \frac{1}{B^{[n]}} \frac{\partial B^{[n]} }{\partial M} \right) 
\neq 0,
\end{eqnarray}
since each term can be written by
\begin{eqnarray}
&& \frac{1}{A^{[n]}} \frac{\partial A^{[n]} }{\partial M} 
= \sum_{p} \frac{1}{M-\omega_{i_p}} + \sum_{p<s<t} \frac{1}{M-\omega_{i_p} - \omega_{i_s} - \omega_{i_t}} + \cdots, \\
&& \frac{1}{B^{[n]}} \frac{\partial B^{[n]} }{\partial M} 
= \frac{1}{M} + \sum_{p<s} \frac{1}{M-\omega_{i_p} - \omega_{i_s}} + \sum_{p<s<t<u} \frac{1}{M-\omega_{i_p} - \omega_{i_s} - \omega_{i_t} - \omega_{i_u} } + \cdots.
\end{eqnarray}
This means that the function ${I^{[n]}}'$ is monotonically increasing or decreasing function. From this fact and \eqref{Relation}, we get the following relation
\begin{eqnarray} 
&& {I^{[n+1]}}' = \frac{  {I^{[n]}}'  }{  {{I}^{[n]}}' \vert_{M \to M - \omega_{i_{n+1}}}   }
 \left\{
\begin{array}{l}
>1 \ ( {I^{[n]}}' : {\rm monotonically \ increasing \ function}  ) \\
<1 \ ( {I^{[n]}}' : {\rm monotonically \ decreasing \ function}  ) 
\end{array}
\right.
.
\end{eqnarray}
Combining with the fact, $\lim_{M \to \infty} {I^{[n]}}' = 1 $ for any $n$, and $ {I^{[1]}}' = (M-\omega_i)/M $, which is a monotonically increasing function, it is easy to show that $ {I^{[n]}}' < 1 $ for odd $n$ and $ {I^{[n]}}' > 1 $ for even $n$. This result shows that $I^{[n]}$ is positive (negative) for even (odd) $n$.

\section{Proof for non-decreasing of $J^{[n]}$}  \label{app2}

In this appendix we show a proof that the total bipartite information \eqref{FunJ} is a decreasing function with respect to the black hole mass $M$. To see this we consider
\begin{eqnarray}\label{eqn:JM_rate}
&&\frac{\partial J^{[n]}}{\partial M}= 8\pi \alpha S(M,\omega_{i_1},\cdots,\omega_{i_n}),
\end{eqnarray}
where we define $S(M,\omega_{i_1},\cdots,\omega_{i_n})$ as
\begin{eqnarray}\label{S}
&&S(M,\omega_{i_1},\cdots,\omega_{i_n}) \equiv \sum_p \frac{1}{M-\omega_{i_p}} -\frac{n-1}{M} -\frac{1}{M- {\displaystyle \sum_p \omega_{i_p} }  }.
\end{eqnarray}
It can be shown that $S$ is a monotonically decreasing function with respect to any $\omega_{i_s}$, that is
\begin{equation}
\frac{\partial S}{\partial \omega_{i_s}} = \frac{1}{(M-\omega_{i_s})^2}-\frac{1}{(M- {\displaystyle \sum_p \omega_{i_p} })^2}=-\frac{  {\displaystyle \sum_{p \neq s}  \omega_{i_p} (M-\omega_{i_s} )  }   }{(M-\omega_{i_s})^2(M- {\displaystyle \sum_p \omega_{i_p} } )^2}<0.
\end{equation}
Therefore, there exists the smallest $\omega=$ min $\{ \omega_{i_p} \}$ such that $S(M,\omega_{i_1},\cdots,\omega_{i_n}) \le S(M,\omega,\cdots,\omega)$.
Using the smallest value of $\omega$, we can check (\ref{eqn:JM_rate}) satisfies the inequality
\begin{equation}
\frac{\partial J^{[n]}}{\partial M} \le 8\pi \alpha S(M,\omega,\cdots,\omega) = -\frac{8\pi \alpha n(n-1) \omega^2 }{M(M-n\omega)(M-\omega)} < 0.
\end{equation}


\end{document}